\documentclass[a4paper,review,11pt,authoryear]{elsarticle}
\usepackage{kpfonts}
\usepackage{amsmath,bm,hyperref,url,fullpage,mathtools,booktabs,parskip}
\usepackage{paralist,enumitem,todonotes,microtype}
\usepackage[a4paper,text={16cm,25cm}]{geometry}
\usepackage[labelsep=period]{caption}
\usepackage[section]{placeins}
\usepackage{algorithm}
\usepackage{algorithmicx}
\usepackage{algpseudocode}
\usepackage{verbatim}
\usepackage{multirow}
\usepackage{multicol}
\usepackage{arydshln}
\usepackage{color}
\usepackage{adjustbox}
\usepackage{amssymb}

\usepackage{hyperref}
\hypersetup{
colorlinks=true,
linkcolor=blue,
citecolor=blue,
urlcolor=blue}

\let\proglang=\textsf
\newcommand{\pkg}[1]{{\fontseries{b}\selectfont #1}}

\journal{}
\begin{document}
\date{}
\begin{frontmatter}

\title{Exploring the social influence of Kaggle virtual community on the M5 competition}

  \author[mainaddress]{Xixi Li\fnref{eqc}}\ead{xixi.li@manchester.ac.uk} \ead[url]{https://orcid.org/0000-0001-5846-3460}
    \author[secondaryaddress]{Yun Bai\fnref{eqc}}\ead{baiyun12138@buaa.edu.cn} 
  \ead[url]{https://orcid.org/0000-0003-4237-7589}
  \author[secondaryaddress]{Yanfei Kang\corref{cor}}\ead{yanfeikang@buaa.edu.cn} \ead[url]{https://orcid.org/0000-0001-8769-6650}
\cortext[cor]{Corresponding author}
  \address[mainaddress]{Department of Mathematics, The University of Manchester, UK.}
  \address[secondaryaddress]{School of Economics and Management,
	Beihang University, China.}
\fntext[eqc]{The authors contributed equally.}

\begin{abstract}
One of the most significant differences of M5 over previous forecasting competitions is that it was held on Kaggle, an online platform of data scientists and machine learning practitioners. Kaggle provides a gathering place, or virtual community, for web users who are interested in the M5 competition. Users can share code, models, features, loss functions, etc. through online notebooks and discussion forums. This paper aims to study the social influence of virtual community on user behaviors in the M5 competition. We first research the content of the M5 virtual community by topic modeling and trend analysis. Further, we perform social media analysis to identify the potential relationship network of the virtual community.  We study the roles and characteristics of some key participants that promote the diffusion of information within the M5 virtual community. Overall, this study provides in-depth insights into the mechanism of the virtual community's influence on the participants and has potential implications for future online competitions.

\end{abstract}


\end{frontmatter}

\newpage

\section{Introduction}
\label{introduction}
The M forecasting competitions \citep{makridakis1982accuracy,makridakis2000m3,makridakis2020m4,makridakis2020m5} were held to identify the most accurate forecasting methods, to advance the theory and practice of forecasting. The task of M5 \citep{makridakis2020m5} is to provide point and interval predictions for each of the 42,840 time series that represent the hierarchical unit sales of the largest retail company in the world, Walmart. Different from the previous offline M competitions, M5 was held on the world's largest online data science community, Kaggle, to attract data scientists and machine learning practitioners to participate.

There is a unique phenomenon that the top methods in the M5 competition are relatively concentrated compared with previous M competitions. For example, a large amount of M5 participants utilize LightGBM \citep{ke2017lightgbm}. Since Kaggle is an online communication platform where participants can produce content and interact with other users \citep{ellison2013sociality}, participants can form virtual communities such as online notebooks and forums to discuss their choices of models, features, loss functions and so on in terms of M5 in the virtual community offered by Kaggle \citep{makridakis2020m5}. 
On the one hand, the virtual community makes it easy for people to interact online with each other. On the other hand, participants' behaviors will be affected by the influential roles in the virtual community, which is called social influence \citep{cialdini1998social}.

To study the social influence of the virtual community on participants in the M5 competition, we adopt a mixed-method approach that combines case studies and social media analysis for several reasons. The mixed-method approach builds on a combination of qualitative and quantitative
methods \citep{aagerfalk2013embracing}. First, the purpose of our research calls for exploratory rather than confirmatory analysis.
The mixed-method approach employs different methods to gain diverse views of the same phenomenon \citep{aagerfalk2013embracing}, which yields additional insights and
enhances the integrity of the findings \citep{creswell2017designing}. Second, it allows us to draw on the strength of the methods used and to offset their weaknesses \citep{creswell2017designing, venkatesh2013bridging}. 
The availability of rich user-generated content (UGC) from the Kaggle platform makes it possible to examine this phenomenon from the perspective of quantitative analysis using text mining and social media analysis. Combined with qualitative analysis like a case study, it can provide deep insights and enhance the understanding derived from the social media analysis \citep{venkatesh2013bridging, ye2021citizen}.

Four crucial factors determine the virtual community's social influence on the participants, namely source, message, channel, and audience \citep{kim2015online}. The credibility and connections with the audience of sources determine the amount of social influence to participants \citep{kim2015online}. Four different types of messages like proprietor contents, user-generated contents, deliberate aggregate user representations, and incidental user representations identified by \cite{walther2012communication} have potential influence on people's behavior. The various applications such as blogs, Twitter, and Facebook may play different roles of channels for social influence. In addition, audience' behavior like sharing, commenting and liking yields social influence \citep{kim2015online}. In this paper, we aim to examine the Kaggle virtual community's social influence on the participants of the M5 competition from the perspective of message and source factors.

The virtual community's contents play an essential role in determining whether social influence has occurred or not. The influence can be examined by the changes in people's feelings, thoughts, behavior, etc. 
The social influence of the M5 virtual community will be reflected upon the choice of methods, features, loss functions, etc., which we aim to associate in this study.
The emergence of large-scale UGC in the M5 virtual community provides a perspective to investigate the association.  Therefore, our first research question (RQ) is as follows.
\begin{itemize}[label={},leftmargin=0.27in]
  \item RQ1:
  \textit{How does UGC from Kaggle M5 Competition virtual community influence participants?}
\end{itemize}

To address this question, we employ topic modeling, an effective method that can help discover the potential topics from large-scale unstructured data. Specifically, we adopt latent Dirichlet allocation (LDA) to discover possible topics from the M5 UGC. We find that these identified topics are centered on LightGBM and Tweedie distribution, which have high consistency with the top solutions in the competition, indicating the social influence of the M5 virtual community on the competition to some extent.
Further, discussions about LightGBM and Tweedie started with posts from influential participants and peaked throughout the period.
We then track the evolution of dynamic topics during the competition with dynamic topic modeling (DTM). This analysis enables us to understand what people were discussing and how the contents changed over time.

Source credibility is conceptualized regarding expertness and trustworthiness in \citep{hovland1953communication}, where they suggest that receivers are willing to accept the information from those who can provide valuable and valid information. Current research like online leaders in the social network is a form of embodiment of source credibility. Online leaders refer to those who can influence others in the community \citep{huffaker2010dimensions}. In this research, we aim to identify some critical roles of the network and explore their abilities in diffusing the LightGBM related information within the network. Our second research question is as follows.
\begin{itemize}[label={},leftmargin=0.27in]
  \item RQ2: 
  \textit{Which roles and characteristics do the key people that promote the diffusion of information within the M5 virtual community have?}
\end{itemize}

To answer this question, we identify some critical roles in the network, including Provider, Supporter, Questioner, Answer person , and Discussion person, and analyze their abilities in spreading information within the network. Each role matters and their cooperation contributes to the information transmission within the social network.

The rest of the article is organized as follows: Section~\ref{method} provides short introductions to the methods that we employ.
Section~\ref{data-analysis} tries to answer the two research questions that we raise from the perspective of topic modeling and social network analysis.
Finally, Section~\ref{conclusion} provides our conclusions.

\section{Methodology}
\label{method}
\subsection{Topic models of documents}
Topic models are a class of statistical models in machine learning that discover abstract topics from documents. The latent Dirichlet allocation (LDA) is one of the most classical topic models using bag-of-words, which treats a document as a collection of words, without order or sequence \citep{blei2003latent}. A document contains multiple topics, to which each word is assigned. LDA can give the topics of each document in the corpus in the form of a probability distribution. For each output topic, some keywords of the document are used for the topic description. LDA has a wide range of applications in text classification~\citep[e.g.,][]{pavlinek2017text}, topic mining~\citep[e.g.,][]{xue2020public}, and forecasting~\citep[e.g.,][]{huberty2015can}. 

One limitation of LDA is the ignorance of the temporal order of documents, especially regarding people's long-term discussions about an event. A dynamic topic model (DTM) designed based on LDA can compensate for this deficiency to some extent \citep{blei2006dynamic}. 
Before implementing DTM, all the documents are divided into slices according to a specific period. Within each slice, the topics of documents are modeled by LDA and considered interchangeable. After that, a normal logistic distribution is applied to describe the evolution of the topics. DTM is more applicable to some documents with time features \citep[e.g.,][]{zhang2015dynamic, jacobi2016quantitative, bai2020topic}.

\subsection{Social network analysis}
Social network analysis that captures the structure of relationships within a network \citep{hoppe2010social}, has been widely used in different areas \citep[e.g.,][]{scott1988social,xixi2017analysis} and recently been applied in the field of social media analysis \citep[e.g.,][]{kwok2018leader}.
The topological analysis aims to study the structural properties of a network \citep{chau2012business}. Some statistics can be employed to quantify the properties of a network. We first present in Table~\ref{tab:key-newtwork-statistics} some key network statistics, including Weighted In-degree (WID), Weighted Out-degree (WOD), Weighted Degree (WD), Closeness Centrality (CC), Betweenness Centrality (BC), and PageRank (PR). 

\begin{table}[!thb]
  \caption{Description of key network statistics.}
\label{tab:key-newtwork-statistics}
\centering
\resizebox{\textwidth}{!}{
\begin{tabular}{p{0.3\columnwidth}p{0.7\columnwidth}}
  \toprule
	Statistic & Description             \\
	\midrule
	Weighted In-degree (WID) & The sum of weighted links to a node.                                                                                       \\
	Weighted Out-degree (WOD) & The sum of weighted links from a node to others.        \\
	Weighted Degree (WD)&The sum of weighted in-degree and weighted out-degree.\\
	Closeness Centrality (CC)&The sum of the distances of one node to all other nodes \citep{golbeck2013analyzing}. It helps us find the individuals who are best placed to influence the entire network most quickly~\citep{Andrew2020}. 
	\\
	Betweenness Centrality (BC)& A measure that captures a person's role in allowing information to pass from one part of the network to another \citep{golbeck2015introduction}.\\
	PageRank (PR)& The importance or influence of a node based on the number and quality of edges to that node \citep{page1999pagerank}.
	\\
  \bottomrule
  \end{tabular}}
\end{table}
\FloatBarrier

A social role is a crucial concept that has been widely studied in different fields, and different methodologies have been employed to investigate social roles within online communities \citep{gleave2009conceptual}. The first conceptual framework for role differentiation comes from the work \citep{gould1989structures}.
In the content of virtual communities, researchers can choose a different perspective to define social roles \citep{benamar2017identification}.
This work aims to study the interactions among the participants and identify some key people that promote the diffusion of information within the M5 virtual community. We provide in Table~\ref{tab:role-definition} the description of some key roles.

\begin{table}[!thb]
  \caption{Description of some key roles.}
\label{tab:role-definition}
\centering
\resizebox{\textwidth}{!}{
\begin{tabular}{p{0.2\columnwidth}p{0.7\columnwidth}}
  \toprule
	Role & Description             \\
	\midrule
	Provider & A Provider is willing to share within the community \citep{fournier2009getting}.                                                                                  \\
  Supporter & A Supporter publishes a large amount of support-related content. They tend to post massive words associated with welcoming new members and devote themselves to solving shared problems \citep{pfeil2011social}.     \\
	Questioner & A Questioner is a person who asks a question within the community, looking for an answer \citep{turner2005picturing}.\\
	Answer person & An Answer person contributes answers within the community \cite{turner2005picturing}.\\
	Discussion person & A Discussion person likes to discussion within the community \citep{fisher2006you}.\\

  \bottomrule
  \end{tabular}}
\end{table}
\FloatBarrier

\section{Data analysis}
\label{data-analysis}
\subsection{M5 UGC acquisition and description}

The online participants' discussion contents studied in this paper are from the accuracy track (concerning point forecasting) of the M5 forecasting competition on Kaggle \footnote{\url{https://www.kaggle.com/c/m5-forecasting-accuracy/discussion}}. 
We employ the \pkg{Selenium}\footnote{\url{https://pypi.org/project/selenium/}} library in \proglang{Python} to crawl a total of 596 posts from the discussion forum, stored as the base corpus for further analysis. 
Post-id posts are numbered sequentially, e.g., Post-18 refers to the 18th post with the title ``First Competition? Say Hi!".
The specific information contains the contents, comments, discussants, and corresponding period of each post. The M5 competition was held on March 3, 2020, with an acceptance deadline of June 24, 2020, and the competition closed on July 1, 2020. The data in this study was collected on January 18, 2021, and the discussions about M5 point forecasting continued from March 3, 2020, to January 12, 2021.

\begin{figure}[!thb]
  \centering
  \includegraphics[width=0.8\textwidth]{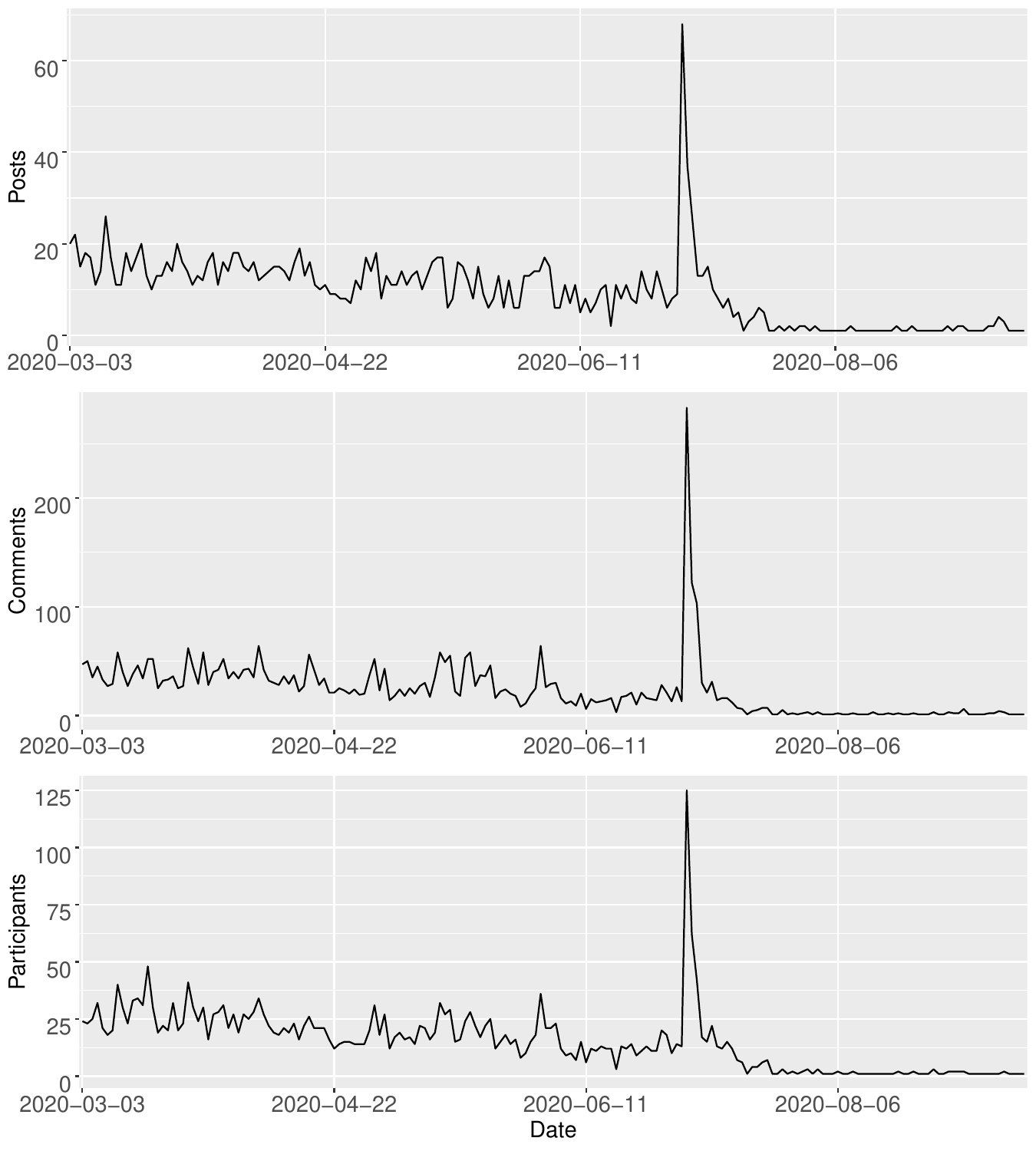}
  \caption{Time series plots of number of posts, comments, and participants.
}
  \label{Description of dataset according to dates.}
\end{figure}

Figure~\ref{Description of dataset according to dates.} shows the number of posts, comments, and active participants in the virtual community per day. It can be observed that these three series have been fluctuating since the competition was released on March 3, 2020, and all reached a maximum peak on July 1, 2020, which is the competition closing date. Since then, the popularity of discussions in the virtual community has declined but continued for several months, indicating that the M5 competition has attracted continuous interest from the community.

\begin{figure}[!thb]
  \centering
  \includegraphics[width=1\textwidth]{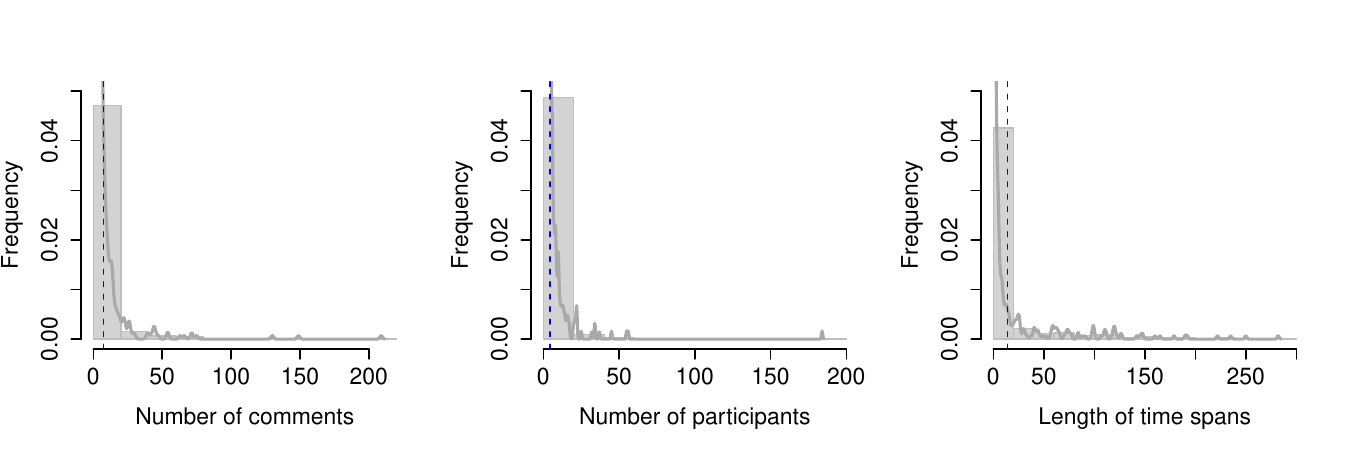}
  \caption{The histograms of the number of comments (left), participants (middle), and time spans (right) of each post.
  }
  \label{Description of dataset according to topics discussed.}
\end{figure}

To have a macro view of the virtual community, we show in Figure~\ref{Description of dataset according to topics discussed.} the histograms superimposed by the corresponding probability density functions in terms of the number of comments, participants and time spans for all posts. 
\emph{Number of comments} refers to the total number of replies posted by the participants under each post. \emph{Number of participants} counts the number of people within the post (i.e., for the case of multiple comments posted by the same person, they are counted only once). \emph{Length of time spans} is the interval of time from when the post was first launched to the latest reply, measured in days. The mean of comments, participants, and the time spans of all posts are 7.41, 4.57, and 13.84, respectively. 45\% of the 596 posts had no more than three comments; 60\% of the posts had less than three participants, and 65\% of the posts lasted no more than three days.
All three subplots in Figure~\ref{Description of dataset according to topics discussed.} exhibit long-tail distributions, with only a pretty small number of extreme values. We show in Table~\ref{Top 5 Posts of Comments, Participants, and TimeSpans} the top five posts in terms of the number of comments, number of participants, and length of time spans. 

\begin{table}[!thb]
  \centering
  \caption{Top five posts in terms of the number of comments, number of participants, and length of time spans. From the first category (comments), Post-18 had the maximum number of comments (209). In terms of participants, Post-18 had the largest number of participants (184). From the last category (Time spans), Post-6 lasted the longest (282 days).}
  \label{Top 5 Posts of Comments, Participants, and TimeSpans}%
  \scalebox{0.8}{
    \begin{tabular}{lllr}
    \toprule
    \textbf{Category} & \textbf{Top Posts} & \textbf{Titles of Posts} & \textbf{Number} \\
    \midrule
    \multirow{5}[2]{*}{Comments} & Post-18    & First Competition? Say Hi! & 209\\
          & Post-13    & Few thoughts about M5 competition & 149\\
          & Post-160    & It's all about upvotes and medals & 130\\
          & Post-167     & Looking for a Team Megathread & 75\\
          & Post-266     & Expressing Frustration: CV vs. LB & 72\\
    \midrule
    \multirow{5}[2]{*}{Participants} & Post-18   & First Competition? Say Hi! & 184  \\
          & Post-167     & Looking for a Team Megathread & 56\\
          & Post-13    & Few thoughts about M5 competition & 55 \\
          & Post-5    & 1st place solution  & 45\\
          & Post-588     & New to Machine Learning or Kaggle? & 37\\
    \midrule
    \multirow{5}[2]{*}{Time spans} & Post-6   & Papers from the M4 Competition & 282 \\
          & Post-7   & Some Objective Function for LightGBM & 250 \\
          & Post-13    & Few thoughts about M5 competition & 235\\
          & Post-18    & First Competition? Say Hi! & 222\\
          & Post-20    & Hierarchical time series in Python & 192\\
    \bottomrule
    \end{tabular}%
    }
  \label{tab:addlabel}%
\end{table}%

\subsection{How does UGC from Kaggle M5 Competition virtual community influence participants?}
\label{topic-modeling}

For the M5 competition discussion forum, the posts with more popularity are worth attention, as shown in Table~\ref{Top 5 Posts of Comments, Participants, and TimeSpans}. 
Nevertheless, the questions and discussions in other posts by the participants contain some specific topics.
Moreover, the knowledge and information containing in the topics may inspire us to the mechanism of how the UGC influences participants' behavior.
While reading and understanding all the posts in detail is undoubtedly a time-consuming task, so we focus on those generalized topics.
Topic modeling is constructed from two aspects, static analysis with LDA model and dynamic analysis with DTM model. Static analysis displays the basic and essential contents in the virtual community. The dynamic analysis further gives more profound insights into how these contents evolve with the social influence of the virtual community.
We first collect all posts at different release times and then employ LDA model to discover the main topics from them, which are static since we ignore the time evolution when extracting topics. Each topic can be described with the corresponding keywords. 
Statistically, these keywords also refer to the words with the highest probability of occurring under each topic and are semantically distinct from those in other topics.
Then topic probability distribution is generated for each post, and the topic with the highest probability is identified as the actual topic of the post. 
After that, to analyze the evolution of the topics, we employ DTM to obtain the evolutionary progress of each specific topic with the dynamic change of topic-word probabilities.

\subsubsection{Static topic analysis of the virtual community}

\begin{figure}[!thb]
  \centering
  \includegraphics[width=1\textwidth]{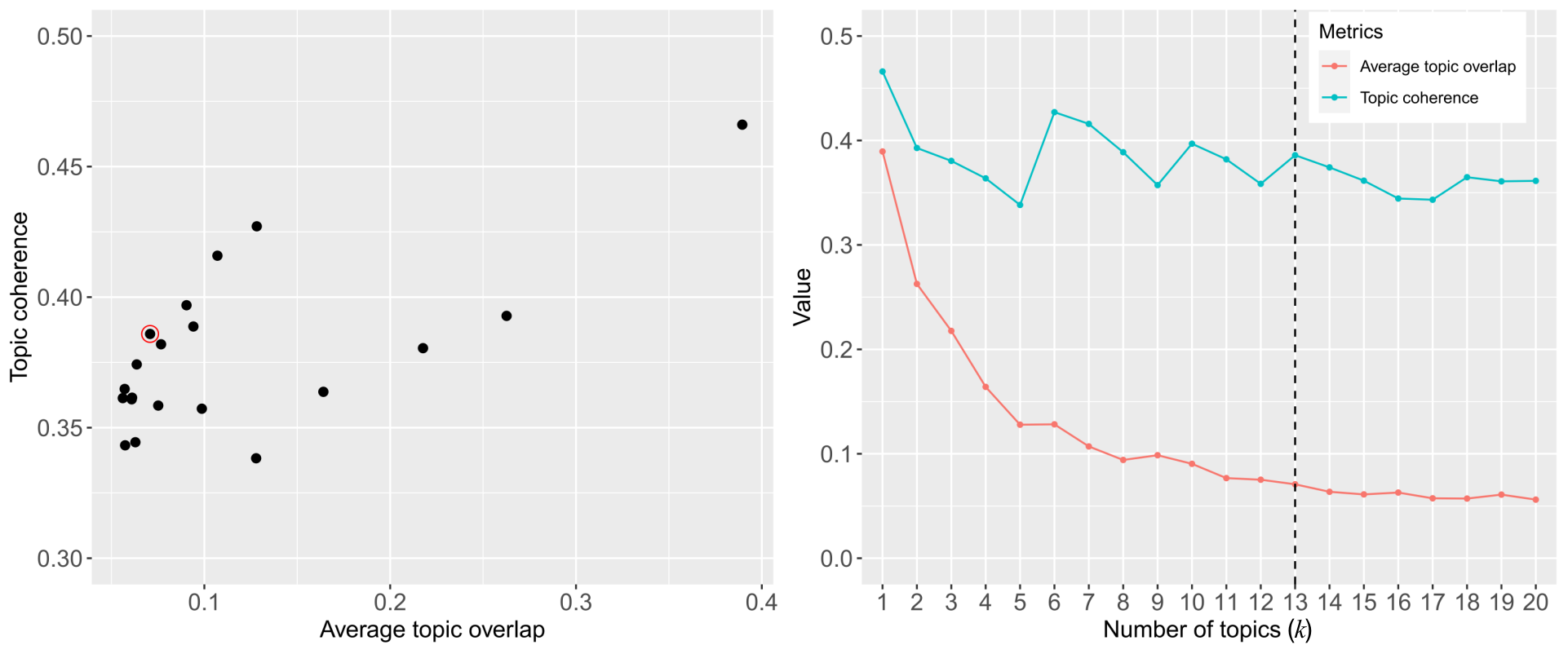}
  \caption{The scatter plot between average topic overlap and topic coherence (\emph{left}). The co-movement of average topic overlap and topic coherence with the number of topics $k$ increasing from one to 20 (\emph{right}). The point circled in red in the scatter plot and  the dashed line in the right panel correspond to the optimal number of topics $k = 13$, when the difference between the two metrics is the largest.}
  \label{num of topics}
\end{figure}

The LDA model requires predetermining the number of topics, denoted as $k$.  To select the optimal number of topics $k$, we consider two metrics named \emph{topic coherence} \citep{newman2010automatic} and \emph{average topic overlap} \citep{o2015analysis}. We calculate the two metrics for $k = 1, 2, \cdots, 20$. From the left panel of Figure~\ref{num of topics}, the topic coherence and average topic overlap are positively correlated. Since a model with higher topic coherence and lower average topic overlap is preferred, we calculate the difference of the two metrics and choose the optimal $k$ when their gap is the largest. The right panel of Figure~\ref{num of topics} shows the co-movement of the two metrics with $k$ increasing from one to 20. The gap between the two lines is the largest when $k = 13$ (shown as the dashed line). Therefore, we use $k = 13$ as the number of topics in our LDA model. The corresponding point is also circled in the left scatter plot of Figure~\ref{num of topics}.

We list the keywords and descriptions under each of the 13 topics detected automatically by LDA in Table~\ref{Key words of 13 topics}.
It can be observed that there are some interesting topics discussed in the M5 virtual community. 
For example, Topic-2 is centered on evaluation; Topic-6 is about some basic description of the dataset. Topic-8 focuses on loss functions, while Topic-10 represents the top solutions (e.g., LightGBM). It is worth mentioning that the Tweedie distribution in Topic-7 and LightGBM in Topic-10 are both leading ideas used in the top methods of the M5 competition.

\begin{table}[htbp]
  \centering
  \caption{Keywords and descriptions of the 13 topics.}
  \label{Key words of 13 topics}%
  \scalebox{0.71}{
    \begin{tabular}{lll}
    \toprule
    \textbf{Topics} & \textbf{Keywords} & \textbf{Summary}\\
    \hline
    Topic-1 & Price, target, rolling, feature, cv, week, store, item, month, encoding & Training Methods\\
    Topic-2 & lb, set, wrmsse, metric, evaluation, period, rmse, training, cv, private & Evaluation\\
    Topic-3 & Error, category, period, understanding, file, forecasts, code, sum, bit, correct & Results submission  \\
    Topic-4 & Dataset, type, ram, memory, results, takes, training, usage, train, predict & Dataset pre-processing \\
    Topic-5 & Private, level, leaderboard, aggregated, lb, unit, total, levels, products, store & Forecasts Ranking \\
    Topic-6 & Product, price, store, week, item, weeks, stores, prediction, demand, historical & Data description\\
    Topic-7 & Recursive, tweedie, level, strategy, loss, distribution, variance, forecasts, store, function & Model training\\
    Topic-8 & Function, custom, loss, objective, wrmsse, notebooks, lightgbm, link, scale, gradient & Loss function\\
    Topic-9 & Solution, team, code, uncertainty, kernel, accuracy, competitions,teams,future,top & Competition rules\\
    Topic-10 & Final, lgbm, regression, set, nn, single, cv, prediction, classfication, based & Top solutions \\
    Topic-11 & Learning, approach, network, training, period, deep, experience, simple, kernels, prediction & Deep learning \\
    Topic-12 & Rows, file, dataset, column, memory, training, test, kernel, evaluation, sample & Data storage \\
    Topic-13 & Level, approach, methods, trend, predictions, space, lstm, modeling, top, neural & Hierarchy forecasting \\
    \hline
    \end{tabular}%
    }
\end{table}%

Although each post has a topic distribution, for simplicity, it is assumed that each post belongs only to the topic with the highest probability. We present the number of posts for each topic in Table~\ref{num of posts under different topics.}, which reflects the popularity of the corresponding topic. It can be seen that Topic-2 (Evaluation) is most popular since it contains the most posts, while Topic-7 (Model training) has the least number of posts. The average number of posts for all topics is 44. 

\begin{table}[hb!]
  \centering
  \caption{Number of posts for each topic. For example, Topic-2 has 77 posts.}
  \label{num of posts under different topics.}
  \scalebox{0.65}{
    \begin{tabular}{lccccccccccccc}
    \toprule
    Topic  & 2 & 6 & 10 & 12 & 4 & 9 & 13 & 11 & 5 & 1 & 8 & 3 & 7 \\ \midrule
    Number of posts   & 77      & 54      & 54       & 52       & 47      & 47      & 45       & 41       & 38      & 36      & 32      & 27      & 23      \\ \bottomrule
    \end{tabular}}
\end{table}

\begin{figure}[!thb]
  \centering
  \includegraphics[width=1\textwidth]{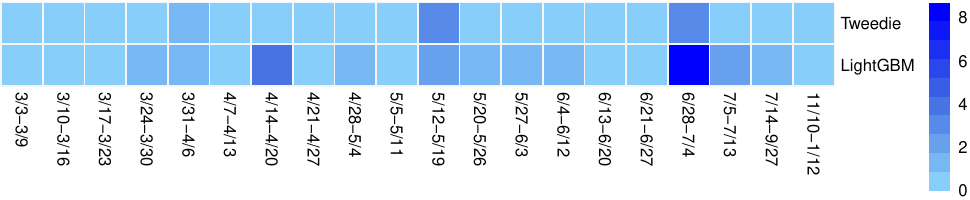}
  \caption{Trends in the discussion about Tweedie and LightGBM. 
  }
  \label{Trends in the discussions about Tweedie and LightGBM}
\end{figure}

Figure~\ref{Trends in the discussions about Tweedie and LightGBM} shows the heatmap of discussions about Tweedie and LightGBM. 
For ease of presentation, we averaged the number of posts discussing Tweedie and LightGBM every seven days. 
Since there is not corresponding discussion every day, the time span of each cell in the heatmap is inconsistent.
In general, the discussions about LightGBM are more heated. 
On April 17, 2020, the post ``Three shades of Dark"\footnote{\url{https://www.kaggle.com/c/m5-forecasting-accuracy/discussion/144067}} was published with sufficient discussions on LightGBM models and training methods. 
The discussions on Tweedie heated on May 13, 2020. A post entitled ``Why Tweedie works?"\footnote{\url{https://www.kaggle.com/c/m5-forecasting-accuracy/discussion/150614}} was published on that day discussing Tweedie regression and distribution, especially on the Tweedie loss function in LightGBM.
Discussions about Tweedie and LightGBM peaked on July 1, 2020, and have been declining ever since.

\subsubsection{Dynamic topic analysis of the virtual community}

\begin{figure}[!thb]
  \centering
  \includegraphics[width=1\textwidth]{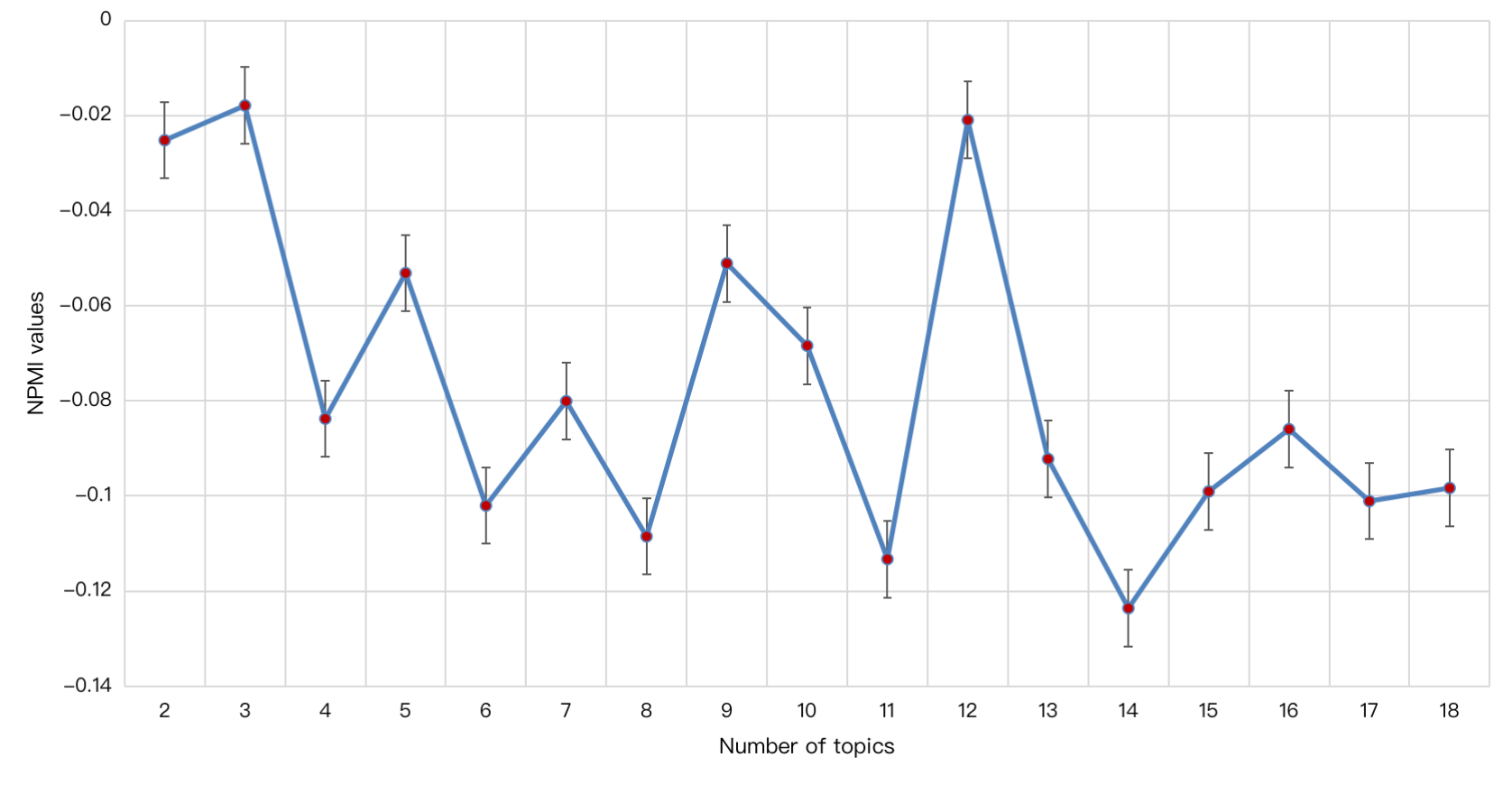}
  \caption{The NPMI values for the different number of topics. 
  }
  \label{The NPMI values of different number of topics}
\end{figure}

In this section, we carry out dynamic topic modeling to the contents of the posts. The total number of valid contents is 573 from 136 days. Considering one month as a time span, we divide the corpus into five slices in chronological order. 
As in \citet{aletras2013evaluating}, normalized pointwise mutual information (NMPI) is deployed to select the appropriate number of topics. The NMPI values with the number of dynamic topics $k$ are shown in Figure~\ref{The NPMI values of different number of topics}, from which it can be viewed that the NPMI value achieves the highest when $k=3$, indicating that the optimal number of topics is three.

\begin{figure}[!thb]
  \centering
  \includegraphics[width=1\textwidth]{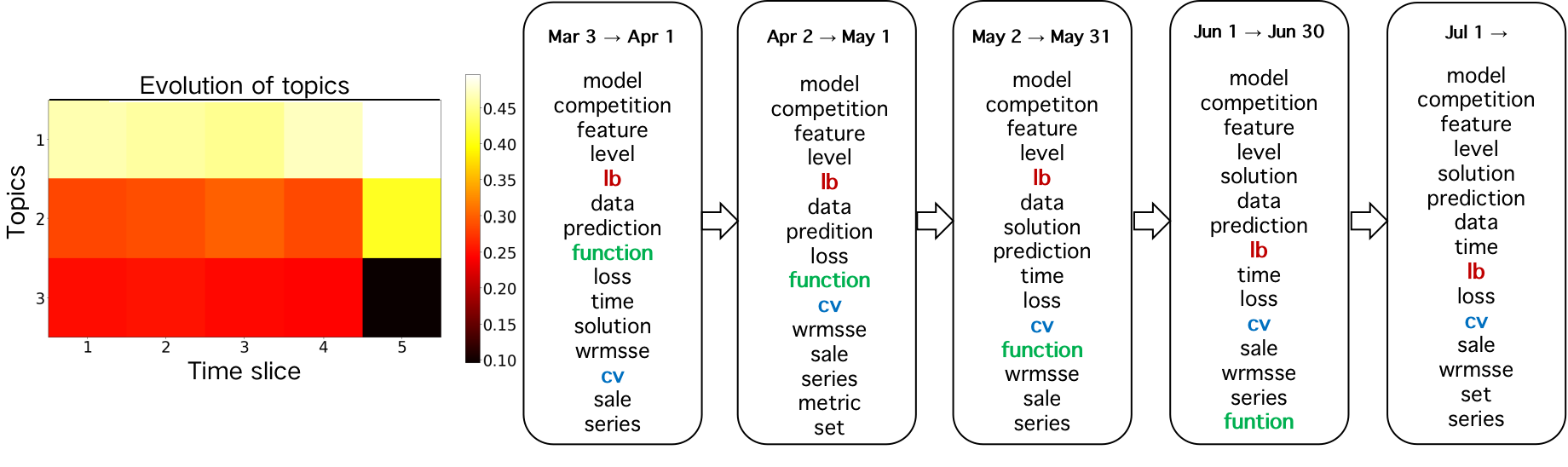}
  \caption{Heatmap of the dynamic topics (\emph{left}). The evolution of keywords in Topic-3 (\emph{right}).
  }
  \label{The evolution of topics and key words}
\end{figure}

We draw a heatmap in the left panel of Figure~\ref{The evolution of topics and key words} to demonstrate the evolution of the dynamic topics over time. 
The lighter the color box in the heatmap is, the more popular the topic at that time slice is. 
As shown in Figure~\ref{The evolution of topics and key words}, Topic-1 and Topic-2 increase while Topic-3 is evolving in the opposite direction.
Taking Topic-3 as an example, we list the corresponding keywords in different time slices, as shown in the right panel of Figure~\ref{The evolution of topics and key words}. `lb' (leaderboard) had been a stable discussion topic in the early stage and gradually started to decline as the M5 competition came to an end; 
the heat of `function' kept falling, and it almost disappeared after July 1.
`cv' (cross validation), used for model training, rose rapidly in popularity a month after the competition started and then began to fluctuate.

Having answered the first question, we begin to examine the phenomenon from social network analysis.

\subsection{Which roles and characteristics do the key people that promote the diffusion of information within the M5 virtual community have?}
\label{social-network-analysis}
This section aims to study the roles and characteristics of some key participants that promote the diffusion of the LightGBM-related information within the M5 virtual community.

\subsubsection{The whole network structure of M5 virtual community}\label{sec:global}
\begin{figure}[!thb]
  \centering
  \includegraphics[width=0.6\textwidth]{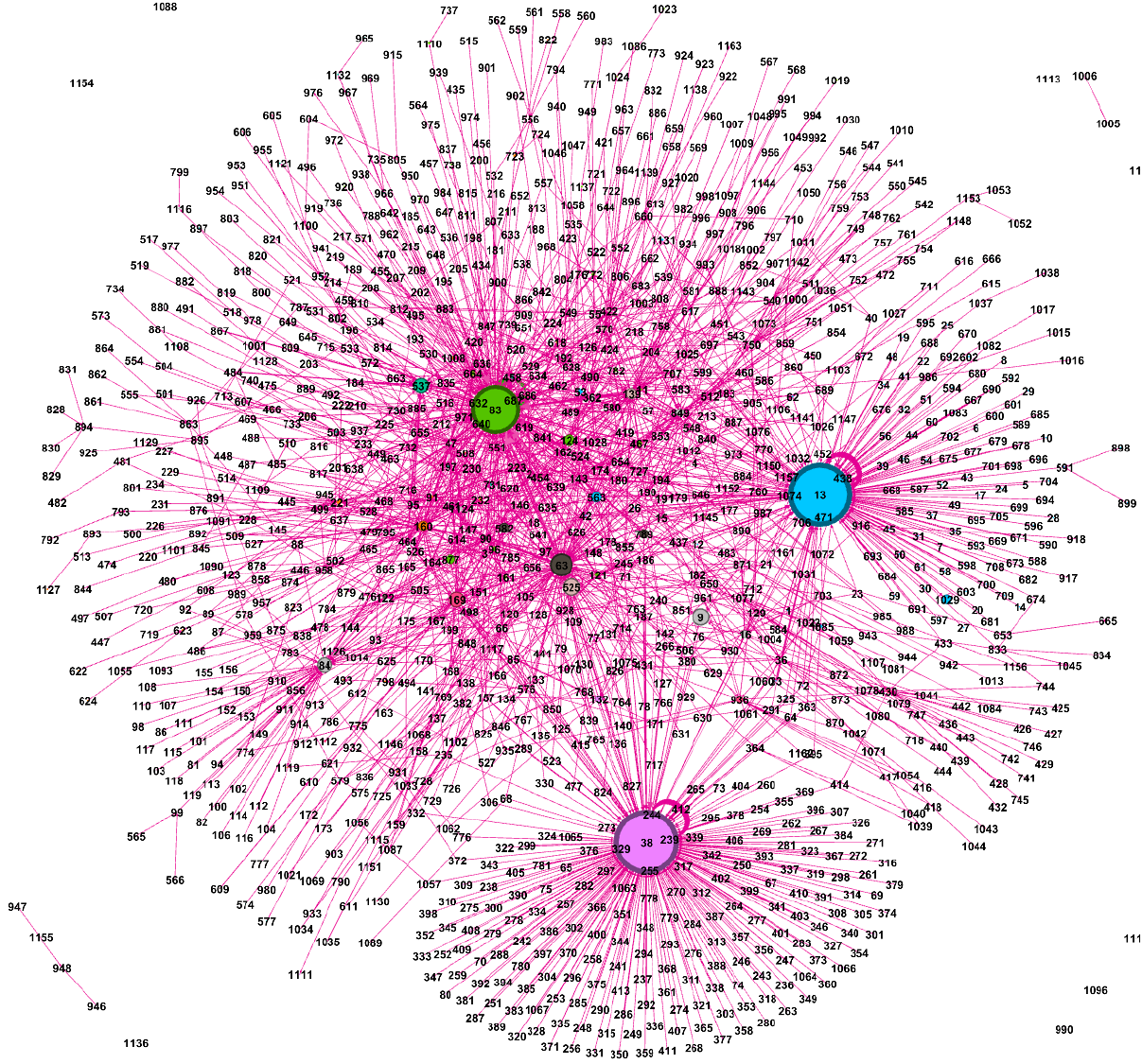}
  \caption{The whole network structure of M5 virtual community.}
  \label{Whole-network}
\end{figure}

The classic network analysis software \proglang{Gephi 0.9.2} \footnote{\url{https://gephi.org}} is used to construct the M5 community network. Figure~\ref{Whole-network} displays the whole structure of the network, where each node represents one participant, and the link between two nodes indicates the interaction between the two participants. These ties are treated as directed and weighted. In other words, when A comments on X's post, then there is a link pointing to X from A. 
The weights refer to the total numbers of comments from one participant to the individual who publishes the posts.
The total numbers of nodes and links of the network are 1165 and 2212, respectively.
The larger the size of a node is, the more in-degree the corresponding participant could be. There exist three supernodes and a large number of ordinary nodes in the network.
Note that some nodes (Id = 13, 38) exhibit self-loops as they comment on their own posts. 28 nodes do not have links with others as they did not receive any replies to their posts.

\subsubsection{Identification and analysis of key roles}
While Section~\ref{sec:global} can be seen as a global interpretation, this section aims to identify some key roles of the networks that promote the diffusion of the LightGBM-related information within the M5 virtual community network. 

WID represents the number of replies a participant received from other participants, which is a crucial indicator of whether the participant's posts are helpful \citep{ye2021citizen}.
The top five participants by WID in the network are shown in Table~\ref{tab:top-5-weighted-in-degree}. Combined with Figure~\ref{Whole-network}, we can observe that these top participants are prominent in the network structure. 
The value in brackets in the PR column in Table~\ref{tab:top-5-weighted-in-degree} and \ref{tab:top-5-wighted-out-degree} represents the ranking of the statistic in the entire community. 
According to PR, we can observe that participants (No.83, No.38, No.13 and No.63) are the four most influential people in the virtual community. 
High BC means that these participants also play a role as a bridge between two other nodes. 
Participants No.63 and No.83 have the highest CC and act as ``broadcasters'' within the network to some extent.

In order to define their roles more accurately, we further track the contents of their posts. It can be observed that the posts of participant No.83 are centered on the topics such as LightGBM-related loss functions, model training, and ensemble models.
Participant No.83 also summarized many top methods about LightGBM and received constant attention and comments, which potentially influenced participants' choices in the M5 competition. Hence, we define this participant as a knowledge provider.
Participants (No.38 and No.13) are Kaggle administrators responsible for posting some notifications about the M5 competition and solving some common problems. They can be viewed as supporters.
Specifically, participant No.38 provided some general strategies about weights, scaling, and aggregation of LightGBM that contributed to problem-solving. Hence, he/she gained wide attention and may influence people's behavior to some extent. 
Participant No.13 acted as a coordinator to help people get started and build teams for the competition.
Participant No.63 is also viewed as a knowledge provider because he/she shared some papers about M4 competitions and hierarchical approaches, which attracted wide attention in the competition. 

\begin{figure}[!thb]
  \centering
  \includegraphics[width=0.6\textwidth]{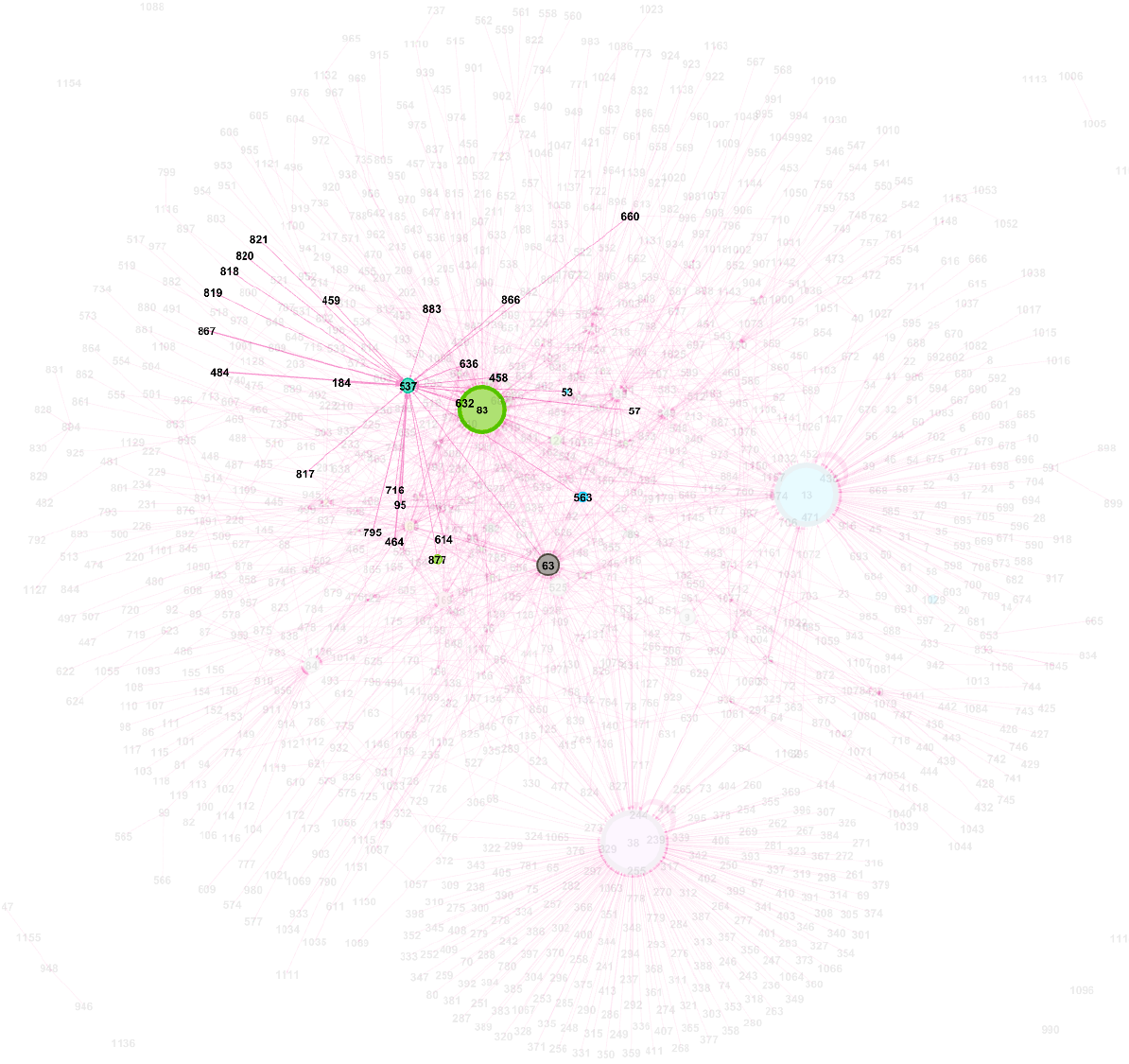}
  \caption{Participant (ID = 537) in the network.}
  \label{Marginal-roles}
\end{figure}

Participant No.537 proposed some general questions about LightGBM, such as cross-validation strategy and loss function problem. These questions are the common problems that everyone will face when using LightGBM, so they received continuous attention. We define this participant as a questioner.
Figure~\ref{Marginal-roles} visualizes the participant (No.537) in the network, who links the central nodes and marginal nodes. In other words, he/she is responsible for spreading information from central participants to marginal participants.

\begin{table}[!thb]
  \caption{Top five participants by WID.}
\label{tab:top-5-weighted-in-degree}
\centering
\resizebox{\textwidth}{!}{
\begin{tabular}{p{0.06\columnwidth}p{0.23\columnwidth}p{0.3\columnwidth}p{0.1\columnwidth}p{0.1\columnwidth}p{0.1\columnwidth}p{0.1\columnwidth}p{0.1\columnwidth}p{0.1\columnwidth}}
  \toprule
	Id&Identity&Role&\textbf{WID}&WOD&WD& CC&BC&PR   \\    	\midrule 
83&Member of Kaggle&Provider&\textbf{488}&150&638&0.3533&81305&0.0204(3)\\
38&Kaggel administrator&Supporter&\textbf{283}&15&298&0.2834&54867&0.0273(1)\\
13&Kaggel administrator&Supporter&\textbf{263}&17&280&0.2565&38616&0.0268(2)\\
63&Member of Kaggle&Provider&\textbf{171}&96&267&0.3863&58390&0.0093(4)\\
537&Member of Kaggle&Questioner&\textbf{90}&35&125&0.2641&5001&0.0059(9)\\

  \end{tabular}}
    \resizebox{\textwidth}{!}{
  \begin{tabular}{p{0.03\columnwidth}p{0.97\columnwidth}}
  \midrule
  \midrule
	Id& Topics of their posts             \\
	\midrule
	83 &High scoring kernels; ensemble and individual models performance; main pipeline; external data; LightGBM base models; cross validation; best single and ensemble score; training time; custom loss and metric.
                                    \\
  38& Public leaderboard; kaggel community; weights, scaling, and aggregation; problem solving.       \\
	13& Top solutions; clarifying questions; external data sources; finding a teammate. 
	\\
		63 &M4 competition related papers; hierarchical approaches; generalization to the private LightGBM; previous time series competitions on Kaggle; WRMSSE metric; public LightGBM; correlation between the local cross validation and the LightGBM.\\
			537 &LightGBM; cross validation; external data; different strategies; score improvement; loss problem; multiplying factor; iterative prediction; categorical feature; KFold.                                                                            \\
		
  \bottomrule
  \end{tabular}}
\end{table}
\FloatBarrier

\begin{figure}[!thb]
  \centering
  \includegraphics[width=0.6\textwidth]{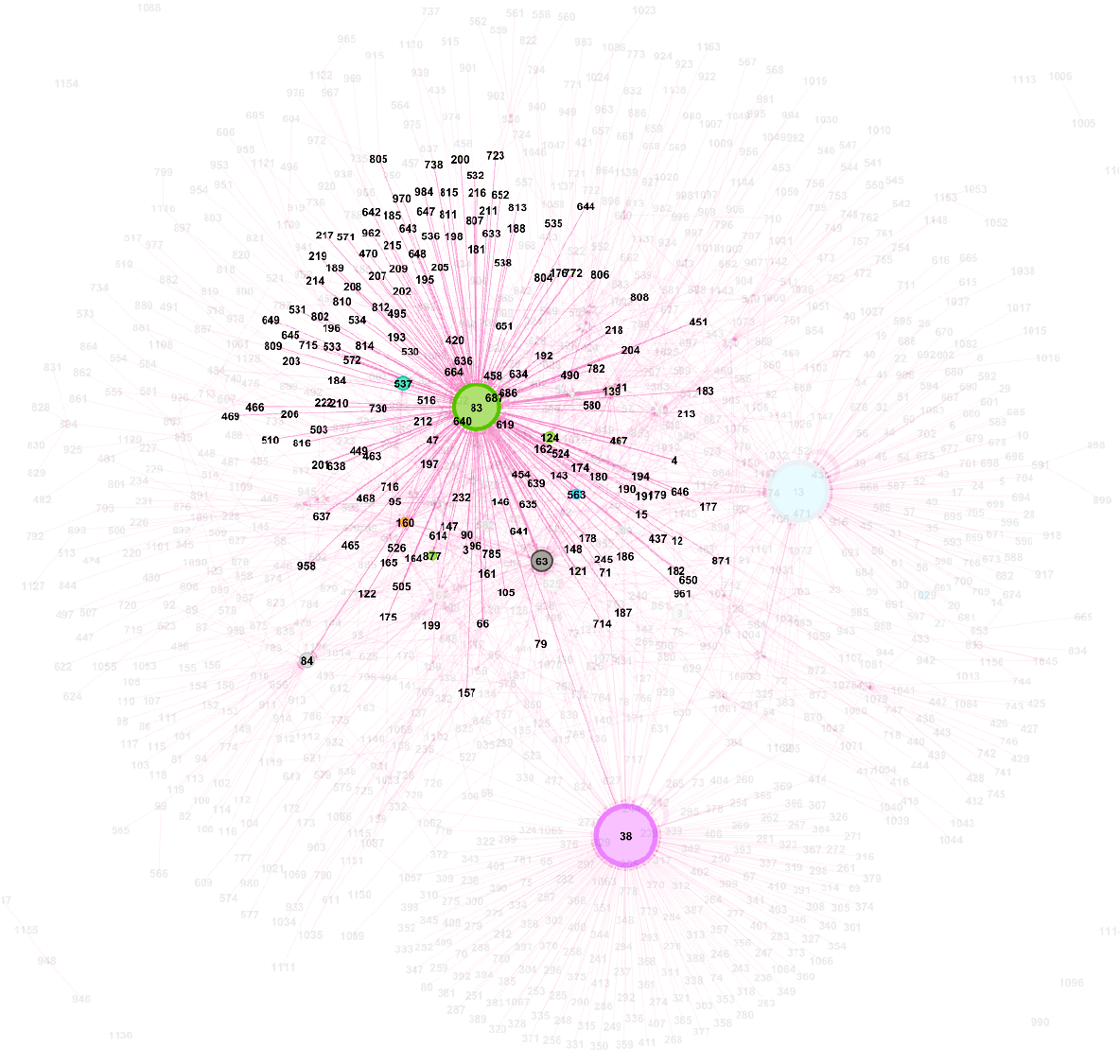}
  \caption{Participant (ID = 83) in the network.}
  \label{Leaders}
\end{figure}

A participant with a high WOD gives many replies to some posts.
The top five participants ordered by this indicator are shown in Table~\ref{tab:top-5-wighted-out-degree}. 
The highest number of responses by one individual in the competition reached 150. We find that participants (No.83 and No.63) also appear in this top five list, indicating their high willingness to provide solutions or ideas to others. 
According to the summary of the replies from participant No.83, he/she played the role of an Answer person and actively offered his/her professional solutions and opinions to others. 
His/Her professional answering focuses more on ensemble models as well as training details of LightGBM. Figure~\ref{Leaders} shows the participant (Id = 83) in the network. Many other participants surround the participant. 
Based on the WID, WOD, and the professions of the No.83 participant's posts, he/she shows leadership for the two aspects that numerous participants follow his/her posts. He/she also actively replied to many others, making contributions to the community.
The leader's influential position in the network makes it possible to influence others once he/she submits a valid and valuable post. 

\begin{figure}[!thb]
  \centering
  \includegraphics[width=0.6\textwidth]{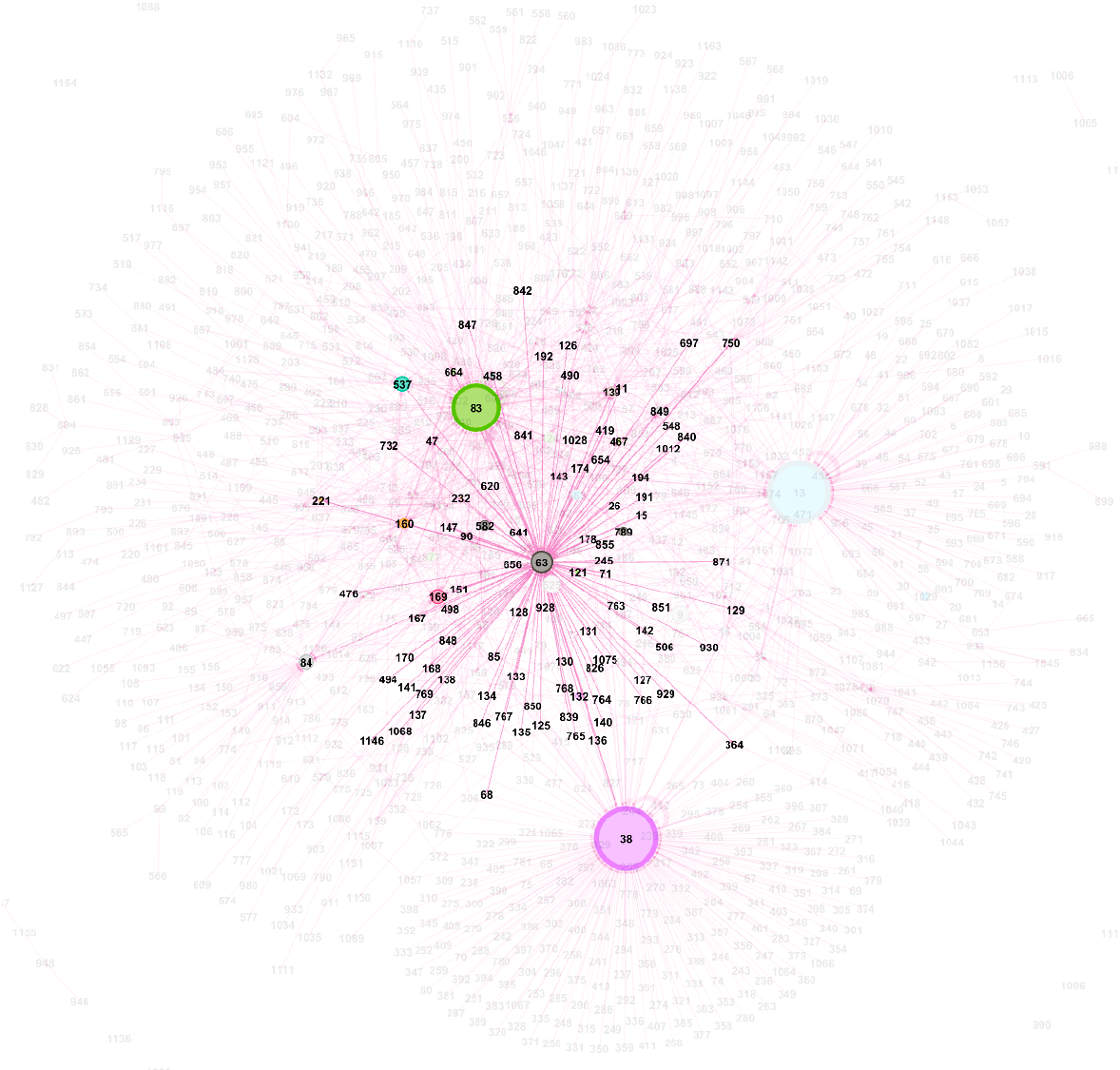}
  \caption{Participant (ID = 63) in the network that links two central nodes.}
  \label{Coordinators}
\end{figure}

While for participant No.63, apart from sharing some knowledge with others, he/she also raised some issues in the community and actively participated in the discussions.
Figure~\ref{Coordinators} shows the participant (Id = 63) in the network, who plays a crucial role in linking different nodes with high WD. He/She also has some followers, and as a result, they may somehow influence others. The participant is essential in a network since he/she makes the information possible to be disseminated through the whole network, which helps avoid network holes. We can observe that participant No.63 focused on raising some issues centered on LightGBM related loss function, parameter tuning, and zero values problem. It can be seen that what he/she posted has some overlaps with the central nodes (participants No.83 and No.38). Also, the followers of participant No.63 have some overlaps with those of the central nodes (participants No.83 and No.38). The two central nodes are linked through these followers, making information spreading possible within the network. 

Participants No.63 and No.47 actively discussed some issues like kernels, LightGBM, and loss functions with others. He/She also raised some questions in the discussions. Participants (No.160 and No.139) are viewed as Answer persons.
Participants No.160 showed his profession in accelerated training and provided his insights to others. For participant No.139, he/she submitted some posts about neural networks and features. We find a few posts related to the neural network based on all the posts,  whose popularity is not as high as we expect compared with those about LightGBM. 

We have made an interactive visualization of the M5 social network, which is publicly available at \url{https://lixixibj.github.io/M5-network-web/}.

\begin{table}[!thb]
  \caption{Top five participants by WOD.}
\label{tab:top-5-wighted-out-degree}
\centering
\resizebox{\textwidth}{!}{
\begin{tabular}{p{0.06\columnwidth}p{0.23\columnwidth}p{0.3\columnwidth}p{0.1\columnwidth}p{0.1\columnwidth}p{0.1\columnwidth}p{0.1\columnwidth}p{0.1\columnwidth}p{0.1\columnwidth}}
  \toprule
	Id&Identity&Role&WID&\textbf{WOD}&WD& CC&BC&PR   \\    	\midrule 
83&Member of Kaggle&Answer person&488&\textbf{150}&638&0.3533&81305&0.0204(3)\\
63&Member of Kaggle&Discussion person&171&\textbf{96}&267&0.3863&58390&0.0093(4)\\
47&Member of Kaggle&Answer person, Discussion person&47&\textbf{93}&140&0.4211&24598&0.0012(54)\\
160&Member of Kaggle&Answer person&69&\textbf{68}&137&0.3473&17238&0.0037(14)\\
	139&Member of Kaggle&Answer person&80&\textbf{61}&141&0.3263&15127&0.0041(11)\\
  \end{tabular}}
    \resizebox{\textwidth}{!}{
  \begin{tabular}{p{0.03\columnwidth}p{0.97\columnwidth}}
  \midrule
  \midrule
	Id& Topics of their replies             \\
	\midrule                                                                           
83&Features; single model; ensemble; regularization; rolling features; Seq2Seq; Tweedie; RMSSE; Training details.\\
63&Problem asking; LightGBM; custom loss function; grid creation; parameter tuning; zero values problem. \\
47&Kernels issue; LightGBM issue; custom loss function; WRMSSE; metric.\\
160&Features; categorical encoding; XGBoost; GPU RAM; post-processing data; speed-up; cross validation; patterns; zero values problem; RNN. \\
			  139& Transformer neural network (NN); ensemble; building a good NN; memory limits; new features.        \\

  \bottomrule
  \end{tabular}}
\end{table}
\FloatBarrier

\FloatBarrier

\section{Conclusion}
\label{conclusion}

This work utilizes advanced text mining technology to study Kaggle virtual community's social influence on participants in the M5 competition, making it possible for people to quickly understand the dynamics of the forum and acquire useful knowledge related to the M5 competition.

We first associate the contents of the M5 virtual community with the competition by topic modeling and trend analysis. We find that posts about competition rules and method summaries tend to have higher level of popularity with respect to the number of comments, participants and time span. 
The topics identified from posts in the Kaggle forum focus on data pre-processing, model training and forecasting evaluation. In particular, there is a lively discussion about novel methods such as Tweedie and LightGBM, which is highly consistent with the final top solutions. In addition, the topics' popularity changes over time.
One of the main contributions of this paper is that several abstract topics are obtained from the vast amount of discussion in the Kaggle forum.
In future, people can then read around these topics. This study gives a direction to researchers to find relevant literature out of an exhaustive list of references.

What's more, we also study the roles and characteristics of the key participants that promoted the diffusion of information within the M5 virtual community.
A relationship network of the M5 virtual community is constructed based on the participants' interactions, with some key roles of the network identified and examined in disseminating the LightGBM-related information within the network. 
The social influence is a relational phenomenon rather than individual status or prestige. 
In the context of the M5 competition, the social influence can come from providing answers to questions, by posting questions in the first place, as well as by being active in disseminating information that has been circulated. 
The identification of these key people can tell a lot about cooperation versus competitive dynamics and can determine how the shared of knowledge happens. Social network analysis provides us with an effective tool to examine these relationships among the participants in the M5 virtual community.

This study opens a novel view on the mechanism of the virtual community’s influence on the competition participants and tries to uncover the hidden connections, patterns, and trends in the virtual community network. 
While having potential implications for future online competitions, this research also has some limitations. 
First, due to the difficulty in acquiring the participants' dynamic decisions of forecasting methods and the related settings such as loss functions and features, we do not quantify the social influence of the M5 virtual community imposed on participants' choices. Instead, we adopt a mixed-method approach that combines case studies and social media analysis. Second, the key roles in the network are identified and defined by combining social network algorithms and the authors' judgmental knowledge. As a result, the social network analysis is influenced by subjective experience to some extent.

\section*{Acknowledgments}
\label{acknowledgements}

The authors are grateful to the editors and two anonymous
reviewers for their helpful comments that improved the contents of the paper. Yanfei Kang is supported by the National Natural Science Foundation of China (No. 72171011 and No. 72021001) and the National Key Research and Development Program (No. 2019YFB1404600).

\bibliographystyle{agsm}
\bibliography{M5-community}
\end{document}